\documentclass[prl,twocolumn,letterpaper,showpacs,groupedaddress,superscriptaddress,longbibliography,floatfix,preprintnumbers,tightenlines]{revtex4-1}
\usepackage[colorlinks,urlcolor=blue]{hyperref}
\usepackage{amsmath,amssymb,amsfonts}
\usepackage{graphicx}
\usepackage{bm}
\usepackage{comment}
\usepackage{color}
\usepackage{csquotes}
\usepackage{longtable}
\usepackage[section]{placeins}

\allowdisplaybreaks
\let\Oldsection\section
\renewcommand{\section}{\FloatBarrier\Oldsection}
\let\Oldsubsection\subsection
\renewcommand{\subsection}{\FloatBarrier\Oldsubsection}

\newcount\hour \newcount\hourminute \newcount\minute
\hour=\time \divide \hour by 60
\hourminute=\hour \multiply \hourminute by 60
\minute=\time \advance \minute by -\hourminute
\newcommand{\mydate}{\ \today \ - \number\hour :\number\minute}
\newcommand{\bea}{\begin{eqnarray}}
\newcommand{\eea}{\end{eqnarray}}

\newcommand{\eq}[1]{Eq.~(\ref{#1})}

\def\g{\gamma}\def\r{\rho}\def\L{\Lambda}\def\a{\alpha}\def\d{\delta}\def\bfb{{\bf b}}\def\z{\zeta}\def\b{\beta}

\def\OMIT#1{{}}

\def\D{\Delta}
\def\bfb{{\bf b}}\def\bfk {{\bf k}}

\def\r{\rho}\def\kp{k_\perp}

\def\G{\Gamma}
\def\vphi{\varphi}
\def\a{\alpha}\def\z{\zeta}

\begin{document}

\title{
Quark  Counting, Drell-Yan West, and  the Pion Wave Function}

\preprint{NT@UW-24-2}
\author{Mary Alberg } 
\affiliation{Department of Physics, Seattle University, Seattle, Washington 98122,USA}
\affiliation{Department of Physics, University of Washington, Seattle, WA 98195-1560, USA}

\author{Gerald A.~Miller }
\affiliation{Department of Physics, University of Washington, Seattle, WA 98195-1560, USA}

\date{\mydate}

\begin{abstract}
  \noindent
  The relation between the pion's quark distribution function, $q(x)$,  its light-front wave function, and the   elastic charge form factor, $F(\D^2)$  is explored.
The square of  the leading-twist pion wave function at a special probe scale, $\z_H$, is determined  using models and Poincare covariance from realistic results for $q(x)$. This wave function is then used to compute form factors with the result  that the Drell-Yan-West  and quark counting relationships are not satisfied. A new relationship between 
$q(x)$ and $F(\D^2)$  is proposed.
  
  \end{abstract}
\maketitle

 \section{Introduction}
 
 The structure of the pion continues to be of interest to many physicists. There are plans to measure the pion electromagnetic form factor at JLab and at the planned EIC~\cite{Aguilar:2019teb}.  There are also plans to re-measure the quark distribution of the pion, $q(x)$  via a new Drell-Yan measurement
 \cite{Adams:2018pwt}.
Much recent and old  theoretical  attention has been devoted to determining and understanding  the behavior of the  valence pion quark distribution function, $q_v(x)$,  at high values of Bjorken $x$, see {\it e.g.} the review~\cite{Holt:2010vj}.   

Much of the recent interest stems from efforts  to understand the behavior at high $x$.
 While many use the parameterization $q_v(x)\sim x^\alpha(1-x)^\beta$ there is a controversy over the value of $\beta$ and its dependence on the variables $x$ and the resolution scale, $Q^2$.
 See, for example,  the differing approaches of~\cite{deTeramond:2018ecg},\cite{Barry:2021osv} and \cite{Cui:2022bxn,Cui:2021mom}.  Ref.~\cite{deTeramond:2018ecg}   finds that $\beta=1$ at low resolution scales, rising to 1.5  at $Q^2=27 $ GeV$^2$, while  \cite{Cui:2021mom} finds that
 $\beta=2+\gamma(Q^2)$ with $\gamma$ positive and increasing at $Q^2$ rises. Both sets of authors claim agreement with the available data set. The small values of
 $\b$ result from perturbative QCD and the larger values from non-perturbative techniques. Indeed,  \cite{Barry:2021osv} finds that  the value of $\b$ can lie between 1 and 2.5 depending on the technique used to resum  the contributions of large logarithms in computing the relationship between $q(x)$ and the measured Drell-Yan cross section data. It would be beneficial to find the relation (if any) between the behavior at large
 values of $x$ and the underlying dynamics. 

The wide interest in the form factor and distribution function originates  in the early hypotheses of the connection between the two observable quantities.
Drell \& Yan~\cite{Drell:1969km} and West~\cite{PhysRevLett.24.1206} suggested a relation between $q(x)$ for large values of $x$ and the elastic proton's Dirac form factor $F_1(\D^2)$, where $\D^2$ is the negative of the square of the four-momentum transfer to the target hadron, at large values of $\D^2$.   Different aspects of the wave function are used to compute distribution functions and form factors, so the following relation is very surprising, namely  
\bea \lim_{x\to1}q(x)=(1-x)^{n_H} \label{DY} \eea
leads to the result
\bea\lim_{\D^2\to\infty} F_1(\D^2)\propto{1\over (\D^2)^{(n_H+1)/2}},\label{DYW}\eea
with $n_H$  the number of partons in the hadron.
The $F_1$ form factor is the matrix element of the plus-component of the electromagnetic current operator between proton states of the same spin. So the  proof would provide the same relations for the pion.
The original papers are heavily quoted now despite the ancient nature of these relations.

Another relationship between structure functions and form factors is obtained from the use of perturbative QCD and leads to quark counting rules for the proton and pion  obtained in ~\cite{Brodsky:1973kr, Brodsky:1974vy, Farrar:1975yb, Farrar:1979aw} and reviewed in~\cite{Sivers:1982wk}. 
These are 
\bea 
\lim_{x\to1} q(x,Q^2)\propto (1-x)^{2n_H-3+2|\D_s|+\D\g},\label{QC}\\
\lim_{\D^2\to\infty} F_H(\D^2)\propto {1\over (\D^2)^{n_H-1}},\label{QC0}\eea
with $ \D\g$ a correction accounting for evolution that vanishes at  a starting scale $\z_H^2$,  
$n_H$ is the minimum number of elementary constituents of the hadron,  and we have taken the number of spectators to be $n_H-1$. 
The quantity $\D_s$ is the difference between the $z$-components of the quark and hadron spin. Thus for a proton the dominant term at high $x$ has $|\D S_z|=0$ and for a pion $|\D S_z|=1/2$. The two sets of relations (\eq{DY}, \eq{DYW}) and (\eq{QC}, \eq{QC0})  are approximately the same for the proton at $\z_H^2$: namely, $q(x)\sim (1-x)^3$ and
$F_1\sim 1/\D^4$  with $n_H=3$.
There are two sets of predictions for the pion $q(x)\sim (1-x)^2, F(\D^2)\sim 1/\D^3$ for  
Drell-Yan West, and $q(x)=(1-x)^2,F(\D^2)\sim 1/\D^2$ for the 
quark counting rules. 
We now focus on the pion.

 The current literature tells us that the relation between the high-$x$ behavior of $q(x)$ and  the pion form factor is interesting.
 In other words, what can the value of $\b$      be used                    for?
We   aim  to study the connection between $q(x)$ and the square of the pion valence light-front  wave function. 
 
\section{Light-Front analysis}
Hadronic wave functions depend on a factorization scale, $\zeta$ at which the hadron is probed.  It has been widely argued that \cite{Jaffe:1980ti,PhysRevD.22.1787,Cui:2021mom} there is a scale at which  the hadron consists of only valence quarks linked to quarks of the quark-parton model as an object dressed by quark-gluon QCD interactions as obtained  from the quark gap equation.  Gluon emission from valence quarks begins at $\z_H$~\cite{Brodsky:1979gy}.  Thus, at $\z_H$  the dressed valence  $u$ and $\bar d$ quarks carry all of the momentum of the $\pi^+$ 
and each constituent (of equal mass) carries 1/2 of the pion momentum.   The result of every calculation of pionic properties that respects Poincare covariance, and the Ward-Green-Takahashi identities along with the consequences of dynamical symmetry breaking inherent in the quark gap-equation has these features. See {\it e.g.} Ref.~\cite{Ding:2019lwe}.

The relation between the  light-front wave function, evaluated at the hadron scale $\zeta_H^2$.
 is given by
\bea q(x)={1\over \pi}\int{d^2k_\perp \over x(1-x)} |\Phi(x,k_\perp)|^2
\label{qdef}
\eea
where $\Phi(x,k_\perp)$ is the wave function of the $q \bar q$ component.  The function $\Phi$ represents  the  leading-twist component of the pion wave function, in which the quark and antiquark spins combine to 0. This component dominates computations of  the high-momentum transfer form factor and the high $x$ behavior of $q(x)$.
The normalization is $\int_0^1dx\,q(x)=1$. We drop the explicit dependence on  $\zeta_H^2$ to simplify the equations.  

There is a special feature of the wave function at $\z_H$.
Rotational invariance   requires that
$\Phi(x,k_\perp)$ is a function of a single variable, $\Phi(M_0^2)$, with   $M_0^2\equiv{k_\perp^2+M^2\over x(1-x)}$,  and $M$ is the constituent quark mass~\cite{osti_7124281,Frankfurt:1981mk}.  The key point is that in the two-body sector one may   construct a self-consistent representation of the Poincare generators. Both $\kp^2 $ and $M^2$ are dimensionless variables measured in terms of an appropriate intrinsic momentum scale, $\L^2$.  Then
changing variables to $z=M_0^2$ leads to the exact result,
\bea q(x)=   \int_{M^2\over x(1-x)}^\infty dz|\Phi(z)|^2\label{q02}.
\eea

A curious feature is that if $M=0$, $q(x)=1$ in disagreement with realistic extractions of $q(x)$ at the hadronic scale~\cite{Cui:2021mom}. Moreover, the idea that there is a scale $\zeta_H$ goes along with the feature that spontaneous symmetry breaking causes $M$ to be significantly larger than its current quark value. Thus we do not expect that $q(x)$ is constant when evaluated at $\zeta_H$.

The next step is  
 to take $x$ to be near unity so that
\bea q_{x\to1}(x)\approx  \int_{M^2\over 1-x}^\infty dz|\Phi(z)|^2.\label{q2}
\eea
The lower limit is large $M^2/(1-x)\gg 1$. This shows immediately the connection between
the large $x$ behavior and the high-momentum part of the light front wave function. 
An interesting relation   can be obtained by differentiating \eq{q2} with respect to $x$:
\bea
q_{x\to1}'(x)=-  {M^2\over (1-x)^2} \Large|\Phi({M^2\over 1-x})\Large|^2.
\label{qp} 
\eea
Given a model wave function one can obtain the high $x$ behavior of $q'(x)$ and thus also that of $q(x)$ at $\zeta^2_H$.  Then DGLAP evolution can be used to obtain the structure functions at larger values of probe scales. \\
Moreover, the finite nature of $q'(x)$ at $x\to1$ immediately gives information about the high momentum behavior of the pion wave function. Namely, 
\bea \lim_{x\to1}   \Large|\Phi({M^2\over 1-x})\Large|^2= c (1-x)^2 f(1-x),\eea where $c$ is a finite number and $f(1-x)$ is finite  as $x\to1$.
Thus the high $x$ behavior of $q(x)$ tells us about specific features of  the pion wave function.
Can one say more?


Form factors are matrix elements of a conserved current and so are  independent of the factorization scale~\cite{Diehl:2003ny} so that one may evaluate the form factor using the  constraint at  $\zeta_H$.  
Then the  form factor is given by the expression
\begin{widetext}
\bea F(\D^2)={1\over \pi}\int_0^1dx \int {d^2k_\perp\over x(1-x) } \Phi({k_\perp^2+M^2\over x(1-x)})\Phi({(\bfk_\perp+(1-x)\boldsymbol{\D})^2+M^2\over x(1-x)}),\label{F}\eea
\end{widetext}
where the plus-component of the  space-like momentum transfer to the proton is taken as zero, so that the momentum transfer, $(\boldsymbol{\D})$  is in a transverse $(\perp$)direction.


To see if there is a connection
  between $F(\D^2)$ and $q(x)$
 we use 
   model wave functions  to compute  both quantities. The connection   between  wave functions and $q(x)$ is given by    \eq{q02}. 
  
 It is convenient to  use   a  flexible power law (PL) form: 
\bea |\Phi^{\rm PL}(z)|^2\to {K\over (z)^{n+1}},q^{\rm PL} \sim (1-x)^n\label{PL}\eea 
with $n\ge 1$. 
This form does not build in the asymptotic behavior predicted by using perturbative QCD. However, the applicability of perturbative QCD to exclusive processes at non-asymptotic, experimentally realizable values of the momentum transfer has been questioned~\cite{Isgur:1988iw,avr1991,Szczepaniak:1997sa,Miller:1994uf,Li:1992nu} for a variety of reasons including lack of knowledge of the non-perturbative part of the wave function, convergence issues,  higher twist effects and those of Sudakov supression. Radyushkin~\cite{,avr1991} wrote, ``for accessible energies and momentum transfers the soft (nonperturbative) contributions dominate over those due to the hard quark rescattering subprocesses". Much of the problems are  related to the importance of the high-$x$ region in computing the form factors that  Feynman argued 
\cite{Feynman1972} was dominant. Despite  progress in  understanding non-perturbative aspects using lattice QCD (see {\it e.g.} \cite{Abdel-Rehim:2015owa}, and Dyson-Schwinger  techniques (see {\it e.g.} ~\cite{Cui:2020tdf,Cui:2021mom}, we believe that it is worthwhile to examine models of non-perturbative wave functions.

With $n=1$, $F(\D^2)\sim 1/\D^3$ with Drell-Yan-West and $F\sim 1/\D^2$ with quark counting. 
These 
predictions can be checked by doing the exact model calculation. We thus expect the asymptotic form factor to behave as  $\sim 1/\D^2$ ,  \eq{QC0}, if   quark counting is correct. 
We now check to see if the quark counting relations are respected if \eq{PL} describes  the wave function.

Note that in the non-relativistic limit that the integral appearing in \eq{F} is dominated by values of $x$ near $1/2$, and if $\D^2\gg (M^2)$  then 
$
 F_{\rm NR}(\D^2)\sim 
 \Phi(1/2\D^2) \sim (1/\D^2)^{(n+1)/2}$ 
  in accord with \eq{DYW}. This result is similar to the non-relativistic arguments presented by Brodsky \& Lepage~\cite{Brodsky:1989pv}.
However, the region of $x$ near unity is very important because the  effects of a large value of $\D$ are mitigated.

To understand this, let's compute the form factor using \eq{PL} in \eq{F} with   $m\equiv (n+1)/2$.
Combining denominators using the Feynman parametrization  and integrating over the transverse momentum variable leads to the result
\bea&
F_m(\D^2)= 
C K_m\int_0^1dx\int_0^1du  { (x(1-x))^{2m-1}(u(1-u))^{m-1}\over (1 +\D^2(1-x)^2 u(1-u))^{2m-1}},\nonumber\\&
\label{FD0}\eea
with $\D^2$ expressed in units of $M^2$, and $CK_m=\frac{\Gamma (4 m) \Gamma (m+1)}{\Gamma (m)^2 \Gamma (2 m)}$. 
A brief look at the integrand of \eq{FD0} shows why it is difficult to determine the asymptotic behavior of $F(\D^2)$. The value of $\D^2$ can be taken to be large, 
but the multiplying factor, $(1-x)^2 u(1-u)$ can be very small. One must do the integral first and then take $\D^2$ to be large.
Closed form expressions for $F_m$  can be obtained for values of $m$ between 1 and 3, 
and
  the asymptotic forms of $F_m$  for $n=1,2,3$  are  shown in Table~I.

\begin{center}
\begin{table}

\begin{tabular}{||c |c||}
\hline
$n$& $\lim_{\D^2\to\infty}F_m(\D^2)$\\
\hline\hline
1 &$6 \left(\frac{\log ^2\left(\Delta ^2\right)-4 \log \left(\Delta ^2\right)+8}{2 \Delta ^2}-\frac{2
   \left(\log \left(\Delta ^2\right)+2\right)}{\Delta ^4}\right)$\\[0.75ex]
 \hline
 2& $180 \sqrt{\pi } \left(\frac{\left(\Delta ^2-6\right) \log \left(\sqrt{\Delta ^2}+\sqrt{\frac{\Delta
   ^2}{4}+1}\right)}{\left(\Delta ^2\right)^{5/2}}+\frac{16-5 \sqrt{\Delta ^2+4}}{2 \Delta ^4}\right)$\\[0.75ex]
   \hline
3&$ 840 \left(\frac{3 \left(\log ^2\left(\Delta ^2\right)-3 \log \left(\Delta ^2\right)+7\right)}{\Delta
   ^6}+\frac{3 \log \left(\Delta ^2\right)-14}{6 \Delta ^4}\right)$\\[0.75ex]
   \hline\hline
 \end{tabular}
 \caption{ Asymptotic behavior of $F_m$. The two leading terms are kept, and $n=2m-1$.}
 \end{table}
  \end{center}  



The results in Table~I and \eq{PL} show  that the Drell-Yan-West relations \eq{DY} and \eq{DYW}  are violated by the logarithms, which are not related to those of perturbative QCD that involve the strong coupling constant $\a_S$.  If one uses the quark counting relations \eq{QC}, \eq{QC0} with $\D\g=0$ from using the hadronic scale, and $n=2n_s, n_H=n_s+1=n/2 +1$ the powers of $\D^2$ do not match.  In particular, if $n=2$ quark counting rules would say $F\sim 1/\D^2$ instead we observe that $F\sim\log\D/\D^3$. Moreover,  the appearance of logarithms in Table~1 shows that the approach to asymptotic limit  is extremely slow. Power law wave functions are not consistent with quark counting rules, but nevertheless are relevant. This is because terms like $\D^2 (1-x)^2 u(1-u)$ appear in  the integrals resulting from the evaluation of Feynman diagrams and the values of $x$ and $u$ approach unity when evaluating integrals.

 {\bf REALISTIC FORM FACTORS}
 
  The next step is to see if the power law form has any phenomenological relevance. To this end, we note that $q(x)$ at
   $\zeta_H^2$ is described as a parameter-free prediction of the pion valence-quark distribution function    in Ref.~\cite{Cui:2020tdf,Cui:2021mom}:
\bea&  q(x) = 375.32 x^2(1 - x)^2\nonumber\\&[1-2.5088 \sqrt{x(1-x)}+2.0250x(1-x)]^2 ,\nonumber\\&
\equiv\sum_{N=4}^8 C_N (x(1-x))^{N/2}.\label{C}\eea
This distribution is defined as   Model 1.
The corresponding pion wave function can then be written in a more general form than \eq{PL} as
\bea \Phi(z)={1\over \sqrt{\pi}}
\sum_{n=3}^5 {A_n\over z^{n/2}  }.
\label{Ph1}\eea
Then,  using \eq{qdef} 
\bea & q(x)=
&= \sum_{n=3}^5 \sum_{N=4}^8 {2\over N}A_nA_{N+2-n}(x(1-x))^{N/2}.
\label{A}\eea
Then  $A_n$ is determined by equating \eq{A} with \eq{C}.
The result is
\bea
q(x)=\sum_{n=3}^5 \sum_{N=4}^8  \widetilde C(n,N+2-n)(x(1-x))^{N/2},
\eea
with $\widetilde C(3,3)=C_4,\,\widetilde C(3,4)=C _5,\,\widetilde C(3,5)=3C_6-(7/4)^2C_7^2/C_8,\,\widetilde C(4,4)=(7/4)^2C_7^2/C_8,\,\widetilde C(4,5)=7/4 C_7,\,\widetilde C(5,5)=4C_8, $ and $\widetilde C(n,m)=\widetilde C(m,n).$

  
  The form factor is obtained from \eq{F} and is given by
  \bea &F(\D^2)=\sum_{m,n=3}^5\widetilde C(n,m)I_{nm}(\D^2)\\&
I_{nm}\equiv {\d_{n+m,N+2}\over B(n/2,m/2)} 
 \int_0^1dx (x(1-x))^{N/2-1}
 \int_0^1du 
 {u^{n/2-1}(1-u)^{m/2-1}\over (1-\D^2(1-x)^2u(1-u))}\nonumber\\&
\label{Fr}  
\eea
  with $B$ the beta function. 
 The results are shown in Fig.~1. The units of $\D^2$ are converted to GeV$^2$ by introducing a mass scale. We use $M=134 $ MeV to reproduce measured data. 
 
 An alternative model, Model 2 is presented in Ref.~\cite{Ding:2019lwe}:
 \bea& \tilde q(x)= 
 213.32 (x (1 - x))^2\nonumber\\&\times(1 - 2.9342 \sqrt{x (1 - x)} + 2.2911 x (1 - x)). 
\label{qt} \eea
 This quark distribution can be rewritten in a form consistent with $\Phi^2(M^2/x(1-x))$:
 \bea \tilde q(x)={C\over (\L^2+{M^2\over x(1-x)})^\a}.
 \label{PLf}\eea
 The  constants are given by ${M^2\over \L^2}=0.0550309$ and $\a=3.26654703$. and  $C$ is for normalization. 
 Note that the end-point  behavior of  the two expressions \eq{qt} and \eq{PLf} are very different with latter $\sim (1-x)^{3.27}$ instead of an exponent of 2. Nevertheless,
 the first 11 moments 
 are reproduced to better than 1\%. The next 5 to better than 2\%. The two distributions are experimentally indistinguishable, showing the elusive behavior of the end-point behavior of $q(x)$.
 
Using \eq{q02} yields the square of the wave function to be 
 \bea \Phi^2(z)={\a C\over \L^{2\a}(1+z)^{1+\a}}.
 \label{Pheq}
 \eea 
 Then using \eq{F}
 the form factor is found to be
  \bea&\tilde F(\D^2)= K\int_0^1 dx \int_0^1du{  (x(1-x))^\a(u(1-u))^{{1\over2}(\a-1)}
 \over [x(1-x) +M^2 +\D^2(1-x)^2u(1-u)]^\a}\nonumber\\&
 \label{TF}
 \eea
 Integration over $u$ leads to the generalized parton distribution $H(x,\D^2)$:
\bea &H(x,\D^2)\propto\nonumber\\&
 \frac{  ((1-x) x)^{2 \beta -1} \, _2F_1\left(\frac{1}{2},2 \beta -1;\beta
   +\frac{1}{2};\frac{(1-x)^2 \Delta ^2}{(1-x)^2 \Delta ^2+4 \left(M^2+(1-x)
   x\right)}\right)}{\left(4 \left(M^2+(1-x)
   x\right)+\Delta ^2 (1-x)^2\right)^{1-2 \beta } }
   \eea
 where $\beta=(1+\a)/2$ and  $_2F_1$ is the hypergeometric function and $H(x,0)=\tilde q(x)$. Both $M^2$ and $\D^2$ are given in units of $\L^2$. We choose $\L^2 =0.36~ \rm GeV^2$ to reproduce data. The corresponds to $M=140 $ MeV.
 
  \begin{figure}[h]
  \centering
 \includegraphics[width=0.4\textwidth]{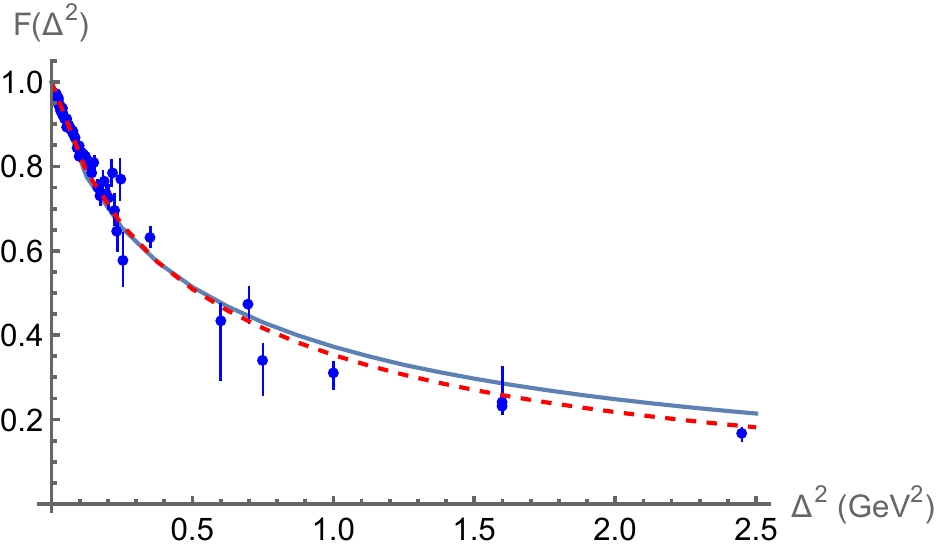}
  \hspace{0.3cm}
  \caption{ $ F(\D^2) $  Solid - Model 1,  $F(\D^2)$ of \eq{F}. Dashed - Model 2, 
  $\tilde F(\D^2)$ of \eq{TF}. The data for $\D^2\le\,0.253 {\rm GeV}^2$ are from CERN Ref.~\cite{NA7:1986vav}. The data for higher values are   from JLab~\cite{JeffersonLab:2008jve}.}
\label{F1} 
 \end{figure}
     \begin{figure}[h]
  \centering
 \includegraphics[width=0.4\textwidth]{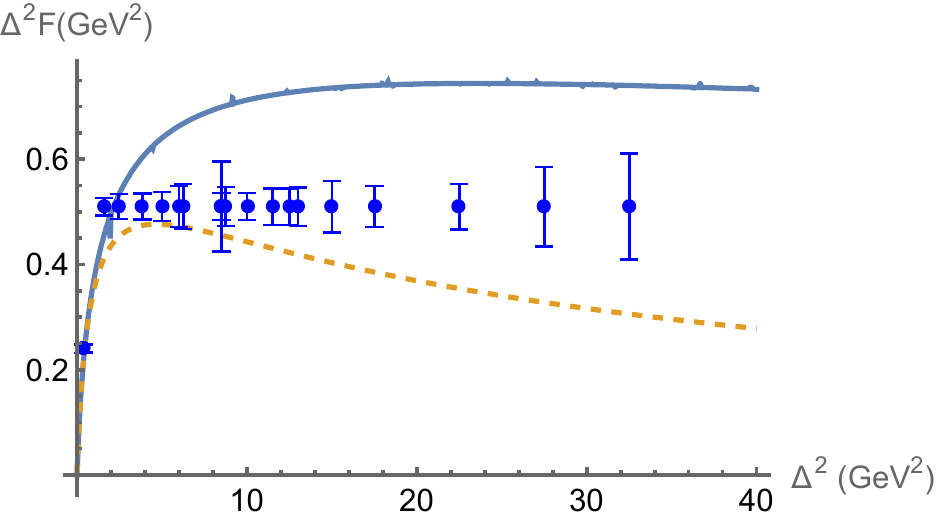}
  \hspace{0.3cm}
  \caption{ $\D^2 F(\D^2) $ in units of GeV$^2$. Solid - Model 1, $F(\D^2)$ of \eq{F}. Dashed - Model 2, 
  $\tilde F(\D^2)$ of \eq{TF}. The projected  error bars for the data points between $\D^2= 0.375$ and 6 Ge V$^2$ are from~\cite{JLab12} and G. M. Huber (private communication).  The projected  error bars for the data points between $\D^2=8.50$  and  15 GeV$2$. are from ~\cite{Accardi:2023chb} and show what might be possible at a 22 GeV facility at JLab. The  projected error bars for higher values of $\D^2$ are from G. M. Huber (private communication) and~\cite{Bylinkin:2022rxd}. In each case the values of $F(\D^2)$ are arbitrary. }
\label{F2} 
 \end{figure}
 
 The result for both form factors are shown in Fig.~\ref{F2}.
There are significant differences in the region that is not yet experimentally explored. 

\medskip
 {\bf CONCLUSIONS}
 
 If the non-perturbative pion wave function can be modeled as a power-law form, or if  the high-$x$  behavior is important for computing the form factor, both the Drell-Yan \& West relations and quark counting rules  for the pion are not correct. Based on our calculations we propose that the
  Drell-Yan-West relations should be modified to
 \bea \lim_{x\to1}q(x)=(1-x)^{n_H} \to  
 F_1(\D^2)\sim{\log(\D^2)\over (\D^2)^{(n_H+1)/2}},\label{AM}\eea
with $q(x)$ evaluated at the hadron scale $\zeta_H^2$, and with the $\log$ not accompanied by a factor involving the strong coupling constant, $\a_S$.

\section*{Acknowledgements}
The work of M.A. is partially supported by the Research in Undergraduate Institutions Program of the US National Science Foundation under Grant No. 2012982.
The work of G.A.M. is partially supported by the DOE Office of Science, Office of Nuclear Physics under Grant  DE-FG02-97ER-41014. 
We thank Garth Huber for providing the data in Figs. 1 \& 2. We thank Craig Roberts for many useful discussions. This work is inspired by the QGT Topical Collaboration.
 \bibliography{miller}

\begin{thebibliography}{35}%
\makeatletter
\providecommand \@ifxundefined [1]{%
 \@ifx{#1\undefined}
}%
\providecommand \@ifnum [1]{%
 \ifnum #1\expandafter \@firstoftwo
 \else \expandafter \@secondoftwo
 \fi
}%
\providecommand \@ifx [1]{%
 \ifx #1\expandafter \@firstoftwo
 \else \expandafter \@secondoftwo
 \fi
}%
\providecommand \natexlab [1]{#1}%
\providecommand \enquote  [1]{``#1''}%
\providecommand \bibnamefont  [1]{#1}%
\providecommand \bibfnamefont [1]{#1}%
\providecommand \citenamefont [1]{#1}%
\providecommand \href@noop [0]{\@secondoftwo}%
\providecommand \href [0]{\begingroup \@sanitize@url \@href}%
\providecommand \@href[1]{\@@startlink{#1}\@@href}%
\providecommand \@@href[1]{\endgroup#1\@@endlink}%
\providecommand \@sanitize@url [0]{\catcode `\\12\catcode `\$12\catcode
  `\&12\catcode `\#12\catcode `\^12\catcode `\_12\catcode `\%12\relax}%
\providecommand \@@startlink[1]{}%
\providecommand \@@endlink[0]{}%
\providecommand \url  [0]{\begingroup\@sanitize@url \@url }%
\providecommand \@url [1]{\endgroup\@href {#1}{\urlprefix }}%
\providecommand \urlprefix  [0]{URL }%
\providecommand \Eprint [0]{\href }%
\providecommand \doibase [0]{http://dx.doi.org/}%
\providecommand \selectlanguage [0]{\@gobble}%
\providecommand \bibinfo  [0]{\@secondoftwo}%
\providecommand \bibfield  [0]{\@secondoftwo}%
\providecommand \translation [1]{[#1]}%
\providecommand \BibitemOpen [0]{}%
\providecommand \bibitemStop [0]{}%
\providecommand \bibitemNoStop [0]{.\EOS\space}%
\providecommand \EOS [0]{\spacefactor3000\relax}%
\providecommand \BibitemShut  [1]{\csname bibitem#1\endcsname}%
\let\auto@bib@innerbib\@empty
\bibitem [{\citenamefont {Aguilar}\ \emph {et~al.}(2019)\citenamefont {Aguilar}
  \emph {et~al.}}]{Aguilar:2019teb}%
  \BibitemOpen
  \bibfield  {author} {\bibinfo {author} {\bibfnamefont {Arlene~C.}\
  \bibnamefont {Aguilar}} \emph {et~al.},\ }\bibfield  {title} {\enquote
  {\bibinfo {title} {{Pion and Kaon Structure at the Electron-Ion Collider}},}\
  }\href {\doibase 10.1140/epja/i2019-12885-0} {\bibfield  {journal} {\bibinfo
  {journal} {Eur. Phys. J. A}\ }\textbf {\bibinfo {volume} {55}},\ \bibinfo
  {pages} {190} (\bibinfo {year} {2019})},\ \Eprint
  {http://arxiv.org/abs/1907.08218} {arXiv:1907.08218 [nucl-ex]} \BibitemShut
  {NoStop}%
\bibitem [{\citenamefont {Adams}\ \emph {et~al.}(2018)\citenamefont {Adams}
  \emph {et~al.}}]{Adams:2018pwt}%
  \BibitemOpen
  \bibfield  {author} {\bibinfo {author} {\bibfnamefont {B.}~\bibnamefont
  {Adams}} \emph {et~al.},\ }\bibfield  {title} {\enquote {\bibinfo {title}
  {{Letter of Intent: A New QCD facility at the M2 beam line of the CERN SPS
  (COMPASS++/AMBER)}},}\ }\href@noop {} {\  (\bibinfo {year} {2018})},\ \Eprint
  {http://arxiv.org/abs/1808.00848} {arXiv:1808.00848 [hep-ex]} \BibitemShut
  {NoStop}%
\bibitem [{\citenamefont {Holt}\ and\ \citenamefont
  {Roberts}(2010)}]{Holt:2010vj}%
  \BibitemOpen
  \bibfield  {author} {\bibinfo {author} {\bibfnamefont {Roy~J.}\ \bibnamefont
  {Holt}}\ and\ \bibinfo {author} {\bibfnamefont {Craig~D.}\ \bibnamefont
  {Roberts}},\ }\bibfield  {title} {\enquote {\bibinfo {title} {{Distribution
  Functions of the Nucleon and Pion in the Valence Region}},}\ }\href {\doibase
  10.1103/RevModPhys.82.2991} {\bibfield  {journal} {\bibinfo  {journal} {Rev.
  Mod. Phys.}\ }\textbf {\bibinfo {volume} {82}},\ \bibinfo {pages}
  {2991--3044} (\bibinfo {year} {2010})},\ \Eprint
  {http://arxiv.org/abs/1002.4666} {arXiv:1002.4666 [nucl-th]} \BibitemShut
  {NoStop}%
\bibitem [{\citenamefont {de~Teramond}\ \emph {et~al.}(2018)\citenamefont
  {de~Teramond}, \citenamefont {Liu}, \citenamefont {Sufian}, \citenamefont
  {Dosch}, \citenamefont {Brodsky},\ and\ \citenamefont
  {Deur}}]{deTeramond:2018ecg}%
  \BibitemOpen
  \bibfield  {author} {\bibinfo {author} {\bibfnamefont {Guy~F.}\ \bibnamefont
  {de~Teramond}}, \bibinfo {author} {\bibfnamefont {Tianbo}\ \bibnamefont
  {Liu}}, \bibinfo {author} {\bibfnamefont {Raza~Sabbir}\ \bibnamefont
  {Sufian}}, \bibinfo {author} {\bibfnamefont {Hans~G\"unter}\ \bibnamefont
  {Dosch}}, \bibinfo {author} {\bibfnamefont {Stanley~J.}\ \bibnamefont
  {Brodsky}}, \ and\ \bibinfo {author} {\bibfnamefont {Alexandre}\ \bibnamefont
  {Deur}} (\bibinfo {collaboration} {HLFHS}),\ }\bibfield  {title} {\enquote
  {\bibinfo {title} {{Universality of Generalized Parton Distributions in
  Light-Front Holographic QCD}},}\ }\href {\doibase
  10.1103/PhysRevLett.120.182001} {\bibfield  {journal} {\bibinfo  {journal}
  {Phys. Rev. Lett.}\ }\textbf {\bibinfo {volume} {120}},\ \bibinfo {pages}
  {182001} (\bibinfo {year} {2018})},\ \Eprint
  {http://arxiv.org/abs/1801.09154} {arXiv:1801.09154 [hep-ph]} \BibitemShut
  {NoStop}%
\bibitem [{\citenamefont {Barry}\ \emph {et~al.}(2021)\citenamefont {Barry},
  \citenamefont {Ji}, \citenamefont {Sato},\ and\ \citenamefont
  {Melnitchouk}}]{Barry:2021osv}%
  \BibitemOpen
  \bibfield  {author} {\bibinfo {author} {\bibfnamefont {P.~C.}\ \bibnamefont
  {Barry}}, \bibinfo {author} {\bibfnamefont {Chueng-Ryong}\ \bibnamefont
  {Ji}}, \bibinfo {author} {\bibfnamefont {N.}~\bibnamefont {Sato}}, \ and\
  \bibinfo {author} {\bibfnamefont {W.}~\bibnamefont {Melnitchouk}} (\bibinfo
  {collaboration} {Jefferson Lab Angular Momentum (JAM)}),\ }\bibfield  {title}
  {\enquote {\bibinfo {title} {{Global QCD Analysis of Pion Parton
  Distributions with Threshold Resummation}},}\ }\href {\doibase
  10.1103/PhysRevLett.127.232001} {\bibfield  {journal} {\bibinfo  {journal}
  {Phys. Rev. Lett.}\ }\textbf {\bibinfo {volume} {127}},\ \bibinfo {pages}
  {232001} (\bibinfo {year} {2021})},\ \Eprint
  {http://arxiv.org/abs/2108.05822} {arXiv:2108.05822 [hep-ph]} \BibitemShut
  {NoStop}%
\bibitem [{\citenamefont {Cui}\ \emph {et~al.}(2022{\natexlab{a}})\citenamefont
  {Cui}, \citenamefont {Ding}, \citenamefont {Morgado}, \citenamefont {Raya},
  \citenamefont {Binosi}, \citenamefont {Chang}, \citenamefont {De~Soto},
  \citenamefont {Roberts}, \citenamefont {Rodr\'\i{}guez-Quintero},\ and\
  \citenamefont {Schmidt}}]{Cui:2022bxn}%
  \BibitemOpen
  \bibfield  {author} {\bibinfo {author} {\bibfnamefont {Z.~F.}\ \bibnamefont
  {Cui}}, \bibinfo {author} {\bibfnamefont {Minghui}\ \bibnamefont {Ding}},
  \bibinfo {author} {\bibfnamefont {J.~M.}\ \bibnamefont {Morgado}}, \bibinfo
  {author} {\bibfnamefont {K.}~\bibnamefont {Raya}}, \bibinfo {author}
  {\bibfnamefont {D.}~\bibnamefont {Binosi}}, \bibinfo {author} {\bibfnamefont
  {L.}~\bibnamefont {Chang}}, \bibinfo {author} {\bibfnamefont
  {F.}~\bibnamefont {De~Soto}}, \bibinfo {author} {\bibfnamefont {C.~D.}\
  \bibnamefont {Roberts}}, \bibinfo {author} {\bibfnamefont {J.}~\bibnamefont
  {Rodr\'\i{}guez-Quintero}}, \ and\ \bibinfo {author} {\bibfnamefont {S.~M.}\
  \bibnamefont {Schmidt}},\ }\bibfield  {title} {\enquote {\bibinfo {title}
  {{Emergence of pion parton distributions}},}\ }\href {\doibase
  10.1103/PhysRevD.105.L091502} {\bibfield  {journal} {\bibinfo  {journal}
  {Phys. Rev. D}\ }\textbf {\bibinfo {volume} {105}},\ \bibinfo {pages}
  {L091502} (\bibinfo {year} {2022}{\natexlab{a}})},\ \Eprint
  {http://arxiv.org/abs/2201.00884} {arXiv:2201.00884 [hep-ph]} \BibitemShut
  {NoStop}%
\bibitem [{\citenamefont {Cui}\ \emph {et~al.}(2022{\natexlab{b}})\citenamefont
  {Cui}, \citenamefont {Ding}, \citenamefont {Morgado}, \citenamefont {Raya},
  \citenamefont {Binosi}, \citenamefont {Chang}, \citenamefont {Papavassiliou},
  \citenamefont {Roberts}, \citenamefont {Rodr\'\i{}guez-Quintero},\ and\
  \citenamefont {Schmidt}}]{Cui:2021mom}%
  \BibitemOpen
  \bibfield  {author} {\bibinfo {author} {\bibfnamefont {Z.~F.}\ \bibnamefont
  {Cui}}, \bibinfo {author} {\bibfnamefont {Minghui}\ \bibnamefont {Ding}},
  \bibinfo {author} {\bibfnamefont {J.~M.}\ \bibnamefont {Morgado}}, \bibinfo
  {author} {\bibfnamefont {K.}~\bibnamefont {Raya}}, \bibinfo {author}
  {\bibfnamefont {D.}~\bibnamefont {Binosi}}, \bibinfo {author} {\bibfnamefont
  {L.}~\bibnamefont {Chang}}, \bibinfo {author} {\bibfnamefont
  {J.}~\bibnamefont {Papavassiliou}}, \bibinfo {author} {\bibfnamefont {C.~D.}\
  \bibnamefont {Roberts}}, \bibinfo {author} {\bibfnamefont {J.}~\bibnamefont
  {Rodr\'\i{}guez-Quintero}}, \ and\ \bibinfo {author} {\bibfnamefont {S.~M.}\
  \bibnamefont {Schmidt}},\ }\bibfield  {title} {\enquote {\bibinfo {title}
  {{Concerning pion parton distributions}},}\ }\href {\doibase
  10.1140/epja/s10050-021-00658-7} {\bibfield  {journal} {\bibinfo  {journal}
  {Eur. Phys. J. A}\ }\textbf {\bibinfo {volume} {58}},\ \bibinfo {pages} {10}
  (\bibinfo {year} {2022}{\natexlab{b}})},\ \Eprint
  {http://arxiv.org/abs/2112.09210} {arXiv:2112.09210 [hep-ph]} \BibitemShut
  {NoStop}%
\bibitem [{\citenamefont {Drell}\ and\ \citenamefont
  {Yan}(1970)}]{Drell:1969km}%
  \BibitemOpen
  \bibfield  {author} {\bibinfo {author} {\bibfnamefont {S.~D.}\ \bibnamefont
  {Drell}}\ and\ \bibinfo {author} {\bibfnamefont {Tung-Mow}\ \bibnamefont
  {Yan}},\ }\bibfield  {title} {\enquote {\bibinfo {title} {{Connection of
  Elastic Electromagnetic Nucleon Form-Factors at Large Q**2 and Deep Inelastic
  Structure Functions Near Threshold}},}\ }\href {\doibase
  10.1103/PhysRevLett.24.181} {\bibfield  {journal} {\bibinfo  {journal} {Phys.
  Rev. Lett.}\ }\textbf {\bibinfo {volume} {24}},\ \bibinfo {pages} {181--185}
  (\bibinfo {year} {1970})}\BibitemShut {NoStop}%
\bibitem [{\citenamefont {West}(1970)}]{PhysRevLett.24.1206}%
  \BibitemOpen
  \bibfield  {author} {\bibinfo {author} {\bibfnamefont {Geoffrey~B.}\
  \bibnamefont {West}},\ }\bibfield  {title} {\enquote {\bibinfo {title}
  {Phenomenological model for the electromagnetic structure of the proton},}\
  }\href {\doibase 10.1103/PhysRevLett.24.1206} {\bibfield  {journal} {\bibinfo
   {journal} {Phys. Rev. Lett.}\ }\textbf {\bibinfo {volume} {24}},\ \bibinfo
  {pages} {1206--1209} (\bibinfo {year} {1970})}\BibitemShut {NoStop}%
\bibitem [{\citenamefont {Brodsky}\ and\ \citenamefont
  {Farrar}(1973)}]{Brodsky:1973kr}%
  \BibitemOpen
  \bibfield  {author} {\bibinfo {author} {\bibfnamefont {Stanley~J.}\
  \bibnamefont {Brodsky}}\ and\ \bibinfo {author} {\bibfnamefont {Glennys~R.}\
  \bibnamefont {Farrar}},\ }\bibfield  {title} {\enquote {\bibinfo {title}
  {{Scaling Laws at Large Transverse Momentum}},}\ }\href {\doibase
  10.1103/PhysRevLett.31.1153} {\bibfield  {journal} {\bibinfo  {journal}
  {Phys. Rev. Lett.}\ }\textbf {\bibinfo {volume} {31}},\ \bibinfo {pages}
  {1153--1156} (\bibinfo {year} {1973})}\BibitemShut {NoStop}%
\bibitem [{\citenamefont {Brodsky}\ and\ \citenamefont
  {Farrar}(1975)}]{Brodsky:1974vy}%
  \BibitemOpen
  \bibfield  {author} {\bibinfo {author} {\bibfnamefont {Stanley~J.}\
  \bibnamefont {Brodsky}}\ and\ \bibinfo {author} {\bibfnamefont {Glennys~R.}\
  \bibnamefont {Farrar}},\ }\bibfield  {title} {\enquote {\bibinfo {title}
  {{Scaling Laws for Large Momentum Transfer Processes}},}\ }\href {\doibase
  10.1103/PhysRevD.11.1309} {\bibfield  {journal} {\bibinfo  {journal} {Phys.
  Rev. D}\ }\textbf {\bibinfo {volume} {11}},\ \bibinfo {pages} {1309}
  (\bibinfo {year} {1975})}\BibitemShut {NoStop}%
\bibitem [{\citenamefont {Farrar}\ and\ \citenamefont
  {Jackson}(1975)}]{Farrar:1975yb}%
  \BibitemOpen
  \bibfield  {author} {\bibinfo {author} {\bibfnamefont {Glennys~R.}\
  \bibnamefont {Farrar}}\ and\ \bibinfo {author} {\bibfnamefont {Darrell~R.}\
  \bibnamefont {Jackson}},\ }\bibfield  {title} {\enquote {\bibinfo {title}
  {{Pion and Nucleon Structure Functions Near x=1}},}\ }\href {\doibase
  10.1103/PhysRevLett.35.1416} {\bibfield  {journal} {\bibinfo  {journal}
  {Phys. Rev. Lett.}\ }\textbf {\bibinfo {volume} {35}},\ \bibinfo {pages}
  {1416} (\bibinfo {year} {1975})}\BibitemShut {NoStop}%
\bibitem [{\citenamefont {Farrar}\ and\ \citenamefont
  {Jackson}(1979)}]{Farrar:1979aw}%
  \BibitemOpen
  \bibfield  {author} {\bibinfo {author} {\bibfnamefont {Glennys~R.}\
  \bibnamefont {Farrar}}\ and\ \bibinfo {author} {\bibfnamefont {Darrell~R.}\
  \bibnamefont {Jackson}},\ }\bibfield  {title} {\enquote {\bibinfo {title}
  {{The Pion Form-Factor}},}\ }\href {\doibase 10.1103/PhysRevLett.43.246}
  {\bibfield  {journal} {\bibinfo  {journal} {Phys. Rev. Lett.}\ }\textbf
  {\bibinfo {volume} {43}},\ \bibinfo {pages} {246} (\bibinfo {year}
  {1979})}\BibitemShut {NoStop}%
\bibitem [{\citenamefont {Sivers}(1982)}]{Sivers:1982wk}%
  \BibitemOpen
  \bibfield  {author} {\bibinfo {author} {\bibfnamefont {Dennis~W.}\
  \bibnamefont {Sivers}},\ }\bibfield  {title} {\enquote {\bibinfo {title}
  {{WHAT CAN WE COUNT ON?}}}\ }\href {\doibase
  10.1146/annurev.ns.32.120182.001053} {\bibfield  {journal} {\bibinfo
  {journal} {Ann. Rev. Nucl. Part. Sci.}\ }\textbf {\bibinfo {volume} {32}},\
  \bibinfo {pages} {149--175} (\bibinfo {year} {1982})}\BibitemShut {NoStop}%
\bibitem [{\citenamefont {Jaffe}\ and\ \citenamefont
  {Ross}(1980)}]{Jaffe:1980ti}%
  \BibitemOpen
  \bibfield  {author} {\bibinfo {author} {\bibfnamefont {R.~L.}\ \bibnamefont
  {Jaffe}}\ and\ \bibinfo {author} {\bibfnamefont {Graham~G.}\ \bibnamefont
  {Ross}},\ }\bibfield  {title} {\enquote {\bibinfo {title} {{Normalizing the
  Renormalization Group Analysis of Deep Inelastic Leptoproduction}},}\ }\href
  {\doibase 10.1016/0370-2693(80)90521-3} {\bibfield  {journal} {\bibinfo
  {journal} {Phys. Lett. B}\ }\textbf {\bibinfo {volume} {93}},\ \bibinfo
  {pages} {313--317} (\bibinfo {year} {1980})}\BibitemShut {NoStop}%
\bibitem [{\citenamefont {De~R\'ujula}\ and\ \citenamefont
  {Martin}(1980)}]{PhysRevD.22.1787}%
  \BibitemOpen
  \bibfield  {author} {\bibinfo {author} {\bibfnamefont {A.}~\bibnamefont
  {De~R\'ujula}}\ and\ \bibinfo {author} {\bibfnamefont {Fran\ifmmode
  \mbox{\c{c}}\else~\c{c}\fi{}ois}\ \bibnamefont {Martin}},\ }\bibfield
  {title} {\enquote {\bibinfo {title} {Structure of hadron structure
  functions},}\ }\href {\doibase 10.1103/PhysRevD.22.1787} {\bibfield
  {journal} {\bibinfo  {journal} {Phys. Rev. D}\ }\textbf {\bibinfo {volume}
  {22}},\ \bibinfo {pages} {1787--1808} (\bibinfo {year} {1980})}\BibitemShut
  {NoStop}%
\bibitem [{\citenamefont {Brodsky}\ and\ \citenamefont
  {Lepage}(1979)}]{Brodsky:1979gy}%
  \BibitemOpen
  \bibfield  {author} {\bibinfo {author} {\bibfnamefont {Stanley~J.}\
  \bibnamefont {Brodsky}}\ and\ \bibinfo {author} {\bibfnamefont {G.~Peter}\
  \bibnamefont {Lepage}},\ }\bibfield  {title} {\enquote {\bibinfo {title}
  {{Perturbative Quantum Chromodynamics}},}\ }\href {\doibase
  10.1007/978-1-4899-6691-9_4} {\bibfield  {journal} {\bibinfo  {journal}
  {Prog. Math. Phys.}\ }\textbf {\bibinfo {volume} {4}},\ \bibinfo {pages}
  {255--422} (\bibinfo {year} {1979})}\BibitemShut {NoStop}%
\bibitem [{\citenamefont {Ding}\ \emph {et~al.}(2020)\citenamefont {Ding},
  \citenamefont {Raya}, \citenamefont {Binosi}, \citenamefont {Chang},
  \citenamefont {Roberts},\ and\ \citenamefont {Schmidt}}]{Ding:2019lwe}%
  \BibitemOpen
  \bibfield  {author} {\bibinfo {author} {\bibfnamefont {Minghui}\ \bibnamefont
  {Ding}}, \bibinfo {author} {\bibfnamefont {Kh\'epani}\ \bibnamefont {Raya}},
  \bibinfo {author} {\bibfnamefont {Daniele}\ \bibnamefont {Binosi}}, \bibinfo
  {author} {\bibfnamefont {Lei}\ \bibnamefont {Chang}}, \bibinfo {author}
  {\bibfnamefont {Craig~D}\ \bibnamefont {Roberts}}, \ and\ \bibinfo {author}
  {\bibfnamefont {Sebastian~M.}\ \bibnamefont {Schmidt}},\ }\bibfield  {title}
  {\enquote {\bibinfo {title} {{Symmetry, symmetry breaking, and pion parton
  distributions}},}\ }\href {\doibase 10.1103/PhysRevD.101.054014} {\bibfield
  {journal} {\bibinfo  {journal} {Phys. Rev. D}\ }\textbf {\bibinfo {volume}
  {101}},\ \bibinfo {pages} {054014} (\bibinfo {year} {2020})},\ \Eprint
  {http://arxiv.org/abs/1905.05208} {arXiv:1905.05208 [nucl-th]} \BibitemShut
  {NoStop}%
\bibitem [{\citenamefont {Terent'ev}(1976)}]{osti_7124281}%
  \BibitemOpen
  \bibfield  {author} {\bibinfo {author} {\bibfnamefont {M~V}\ \bibnamefont
  {Terent'ev}},\ }\bibfield  {title} {\enquote {\bibinfo {title} {On the
  structure of the wave functions of mesons considered as bound states of
  relativistic quarks},}\ }\href {https://www.osti.gov/biblio/7124281}
  {\bibfield  {journal} {\bibinfo  {journal} {Sov. J. Nucl. Phys. (Engl.
  Transl.); (United States)}\ }\textbf {\bibinfo {volume} {24}},\ \bibinfo
  {pages} {1} (\bibinfo {year} {1976})}\BibitemShut {NoStop}%
\bibitem [{\citenamefont {Frankfurt}\ and\ \citenamefont
  {Strikman}(1981)}]{Frankfurt:1981mk}%
  \BibitemOpen
  \bibfield  {author} {\bibinfo {author} {\bibfnamefont {L.~L.}\ \bibnamefont
  {Frankfurt}}\ and\ \bibinfo {author} {\bibfnamefont {M.~I.}\ \bibnamefont
  {Strikman}},\ }\bibfield  {title} {\enquote {\bibinfo {title} {{High-Energy
  Phenomena, Short Range Nuclear Structure and QCD}},}\ }\href {\doibase
  10.1016/0370-1573(81)90129-0} {\bibfield  {journal} {\bibinfo  {journal}
  {Phys. Rept.}\ }\textbf {\bibinfo {volume} {76}},\ \bibinfo {pages}
  {215--347} (\bibinfo {year} {1981})}\BibitemShut {NoStop}%
\bibitem [{\citenamefont {Diehl}(2003)}]{Diehl:2003ny}%
  \BibitemOpen
  \bibfield  {author} {\bibinfo {author} {\bibfnamefont {M.}~\bibnamefont
  {Diehl}},\ }\bibfield  {title} {\enquote {\bibinfo {title} {{Generalized
  parton distributions}},}\ }\href {\doibase 10.1016/j.physrep.2003.08.002}
  {\bibfield  {journal} {\bibinfo  {journal} {Phys. Rept.}\ }\textbf {\bibinfo
  {volume} {388}},\ \bibinfo {pages} {41--277} (\bibinfo {year} {2003})},\
  \Eprint {http://arxiv.org/abs/hep-ph/0307382} {arXiv:hep-ph/0307382}
  \BibitemShut {NoStop}%
\bibitem [{\citenamefont {Isgur}\ and\ \citenamefont
  {Llewellyn~Smith}(1989)}]{Isgur:1988iw}%
  \BibitemOpen
  \bibfield  {author} {\bibinfo {author} {\bibfnamefont {Nathan}\ \bibnamefont
  {Isgur}}\ and\ \bibinfo {author} {\bibfnamefont {C.~H.}\ \bibnamefont
  {Llewellyn~Smith}},\ }\bibfield  {title} {\enquote {\bibinfo {title} {{The
  Applicability of Perturbative QCD to Exclusive Processes}},}\ }\href
  {\doibase 10.1016/0550-3213(89)90532-4} {\bibfield  {journal} {\bibinfo
  {journal} {Nucl. Phys. B}\ }\textbf {\bibinfo {volume} {317}},\ \bibinfo
  {pages} {526--572} (\bibinfo {year} {1989})}\BibitemShut {NoStop}%
\bibitem [{\citenamefont {Radyushkin}(1991)}]{avr1991}%
  \BibitemOpen
  \bibfield  {author} {\bibinfo {author} {\bibfnamefont {A.V.}\ \bibnamefont
  {Radyushkin}},\ }\bibfield  {title} {\enquote {\bibinfo {title} {Hadronic
  form factors: Perturbative qcd vs qcd sum rules},}\ }\href {\doibase
  https://doi.org/10.1016/0375-9474(91)90691-X} {\bibfield  {journal} {\bibinfo
   {journal} {Nuclear Physics A}\ }\textbf {\bibinfo {volume} {532}},\ \bibinfo
  {pages} {141--154} (\bibinfo {year} {1991})}\BibitemShut {NoStop}%
\bibitem [{\citenamefont {Szczepaniak}\ \emph {et~al.}(1998)\citenamefont
  {Szczepaniak}, \citenamefont {Radyushkin},\ and\ \citenamefont
  {Ji}}]{Szczepaniak:1997sa}%
  \BibitemOpen
  \bibfield  {author} {\bibinfo {author} {\bibfnamefont {Adam}\ \bibnamefont
  {Szczepaniak}}, \bibinfo {author} {\bibfnamefont {Anatoly}\ \bibnamefont
  {Radyushkin}}, \ and\ \bibinfo {author} {\bibfnamefont {Chueng-Ryong}\
  \bibnamefont {Ji}},\ }\bibfield  {title} {\enquote {\bibinfo {title}
  {{Consistent analysis of O (alpha-s) corrections to pion elastic
  form-factor}},}\ }\href {\doibase 10.1103/PhysRevD.57.2813} {\bibfield
  {journal} {\bibinfo  {journal} {Phys. Rev. D}\ }\textbf {\bibinfo {volume}
  {57}},\ \bibinfo {pages} {2813--2822} (\bibinfo {year} {1998})},\ \Eprint
  {http://arxiv.org/abs/hep-ph/9708237} {arXiv:hep-ph/9708237} \BibitemShut
  {NoStop}%
\bibitem [{\citenamefont {Miller}\ and\ \citenamefont
  {Pasupathy}(1994)}]{Miller:1994uf}%
  \BibitemOpen
  \bibfield  {author} {\bibinfo {author} {\bibfnamefont {G.~A.}\ \bibnamefont
  {Miller}}\ and\ \bibinfo {author} {\bibfnamefont {J.}~\bibnamefont
  {Pasupathy}},\ }\bibfield  {title} {\enquote {\bibinfo {title} {{Higher twist
  and higher order contributions to the pion electromagnetic form-factor}},}\
  }\href {\doibase 10.1007/BF01289600} {\bibfield  {journal} {\bibinfo
  {journal} {Z. Phys. A}\ }\textbf {\bibinfo {volume} {348}},\ \bibinfo {pages}
  {123--127} (\bibinfo {year} {1994})}\BibitemShut {NoStop}%
\bibitem [{\citenamefont {Li}\ and\ \citenamefont {Sterman}(1992)}]{Li:1992nu}%
  \BibitemOpen
  \bibfield  {author} {\bibinfo {author} {\bibfnamefont {Hsiang-nan}\
  \bibnamefont {Li}}\ and\ \bibinfo {author} {\bibfnamefont {George~F.}\
  \bibnamefont {Sterman}},\ }\bibfield  {title} {\enquote {\bibinfo {title}
  {{The Perturbative pion form-factor with Sudakov suppression}},}\ }\href
  {\doibase 10.1016/0550-3213(92)90643-P} {\bibfield  {journal} {\bibinfo
  {journal} {Nucl. Phys. B}\ }\textbf {\bibinfo {volume} {381}},\ \bibinfo
  {pages} {129--140} (\bibinfo {year} {1992})}\BibitemShut {NoStop}%
\bibitem [{\citenamefont {Feynman}(1972)}]{Feynman1972}%
  \BibitemOpen
  \bibfield  {author} {\bibinfo {author} {\bibfnamefont {Richard~P.}\
  \bibnamefont {Feynman}},\ }\href@noop {} {\emph {\bibinfo {title}
  {Photon-Hadron Interactions}}}\ (\bibinfo  {publisher} {W. A. Benjamin,
  Inc.},\ \bibinfo {address} {Reading},\ \bibinfo {year} {1972})\BibitemShut
  {NoStop}%
\bibitem [{\citenamefont {Abdel-Rehim}\ \emph {et~al.}(2015)\citenamefont
  {Abdel-Rehim} \emph {et~al.}}]{Abdel-Rehim:2015owa}%
  \BibitemOpen
  \bibfield  {author} {\bibinfo {author} {\bibfnamefont {A.}~\bibnamefont
  {Abdel-Rehim}} \emph {et~al.},\ }\bibfield  {title} {\enquote {\bibinfo
  {title} {{Nucleon and pion structure with lattice QCD simulations at physical
  value of the pion mass}},}\ }\href {\doibase 10.1103/PhysRevD.92.114513}
  {\bibfield  {journal} {\bibinfo  {journal} {Phys. Rev. D}\ }\textbf {\bibinfo
  {volume} {92}},\ \bibinfo {pages} {114513} (\bibinfo {year} {2015})},\
  \bibinfo {note} {[Erratum: Phys.Rev.D 93, 039904 (2016)]},\ \Eprint
  {http://arxiv.org/abs/1507.04936} {arXiv:1507.04936 [hep-lat]} \BibitemShut
  {NoStop}%
\bibitem [{\citenamefont {Cui}\ \emph {et~al.}(2020)\citenamefont {Cui},
  \citenamefont {Ding}, \citenamefont {Gao}, \citenamefont {Raya},
  \citenamefont {Binosi}, \citenamefont {Chang}, \citenamefont {Roberts},
  \citenamefont {Rodr\'\i{}guez-Quintero},\ and\ \citenamefont
  {Schmidt}}]{Cui:2020tdf}%
  \BibitemOpen
  \bibfield  {author} {\bibinfo {author} {\bibfnamefont {Zhu-Fang}\
  \bibnamefont {Cui}}, \bibinfo {author} {\bibfnamefont {Minghui}\ \bibnamefont
  {Ding}}, \bibinfo {author} {\bibfnamefont {Fei}\ \bibnamefont {Gao}},
  \bibinfo {author} {\bibfnamefont {Kh\'epani}\ \bibnamefont {Raya}}, \bibinfo
  {author} {\bibfnamefont {Daniele}\ \bibnamefont {Binosi}}, \bibinfo {author}
  {\bibfnamefont {Lei}\ \bibnamefont {Chang}}, \bibinfo {author} {\bibfnamefont
  {Craig~D}\ \bibnamefont {Roberts}}, \bibinfo {author} {\bibfnamefont {Jose}\
  \bibnamefont {Rodr\'\i{}guez-Quintero}}, \ and\ \bibinfo {author}
  {\bibfnamefont {Sebastian~M}\ \bibnamefont {Schmidt}},\ }\bibfield  {title}
  {\enquote {\bibinfo {title} {{Kaon and pion parton distributions}},}\ }\href
  {\doibase 10.1140/epjc/s10052-020-08578-4} {\bibfield  {journal} {\bibinfo
  {journal} {Eur. Phys. J. C}\ }\textbf {\bibinfo {volume} {80}},\ \bibinfo
  {pages} {1064} (\bibinfo {year} {2020})}\BibitemShut {NoStop}%
\bibitem [{\citenamefont {Brodsky}\ and\ \citenamefont
  {Lepage}(1989)}]{Brodsky:1989pv}%
  \BibitemOpen
  \bibfield  {author} {\bibinfo {author} {\bibfnamefont {Stanley~J.}\
  \bibnamefont {Brodsky}}\ and\ \bibinfo {author} {\bibfnamefont {G.~Peter}\
  \bibnamefont {Lepage}},\ }\bibfield  {title} {\enquote {\bibinfo {title}
  {{Exclusive Processes in Quantum Chromodynamics}},}\ }\href {\doibase
  10.1142/9789814503266_0002} {\bibfield  {journal} {\bibinfo  {journal} {Adv.
  Ser. Direct. High Energy Phys.}\ }\textbf {\bibinfo {volume} {5}},\ \bibinfo
  {pages} {93--240} (\bibinfo {year} {1989})}\BibitemShut {NoStop}%
\bibitem [{\citenamefont {Amendolia}\ \emph {et~al.}(1986)\citenamefont
  {Amendolia} \emph {et~al.}}]{NA7:1986vav}%
  \BibitemOpen
  \bibfield  {author} {\bibinfo {author} {\bibfnamefont {S.~R.}\ \bibnamefont
  {Amendolia}} \emph {et~al.} (\bibinfo {collaboration} {NA7}),\ }\bibfield
  {title} {\enquote {\bibinfo {title} {{A Measurement of the Space - Like Pion
  Electromagnetic Form-Factor}},}\ }\href {\doibase
  10.1016/0550-3213(86)90437-2} {\bibfield  {journal} {\bibinfo  {journal}
  {Nucl. Phys. B}\ }\textbf {\bibinfo {volume} {277}},\ \bibinfo {pages} {168}
  (\bibinfo {year} {1986})}\BibitemShut {NoStop}%
\bibitem [{\citenamefont {Huber}\ \emph {et~al.}(2008)\citenamefont {Huber}
  \emph {et~al.}}]{JeffersonLab:2008jve}%
  \BibitemOpen
  \bibfield  {author} {\bibinfo {author} {\bibfnamefont {G.~M.}\ \bibnamefont
  {Huber}} \emph {et~al.} (\bibinfo {collaboration} {Jefferson Lab}),\
  }\bibfield  {title} {\enquote {\bibinfo {title} {{Charged pion form-factor
  between Q**2 = 0.60-GeV**2 and 2.45-GeV**2. II. Determination of, and results
  for, the pion form-factor}},}\ }\href {\doibase 10.1103/PhysRevC.78.045203}
  {\bibfield  {journal} {\bibinfo  {journal} {Phys. Rev. C}\ }\textbf {\bibinfo
  {volume} {78}},\ \bibinfo {pages} {045203} (\bibinfo {year} {2008})},\
  \Eprint {http://arxiv.org/abs/0809.3052} {arXiv:0809.3052 [nucl-ex]}
  \BibitemShut {NoStop}%
\bibitem [{JLa()}]{JLab12}%
  \BibitemOpen
  \href@noop {} {\enquote {\bibinfo {title} {{Measurement of the Charged Pion
  Form Factor to High $Q^2$ }},}\ }\bibinfo {howpublished}
  {\url{https://www.jlab.org/exp_prog/proposals/06/PR12-06-101.pdf}}\BibitemShut
  {NoStop}%
\bibitem [{\citenamefont {Accardi}\ \emph {et~al.}(2023)\citenamefont {Accardi}
  \emph {et~al.}}]{Accardi:2023chb}%
  \BibitemOpen
  \bibfield  {author} {\bibinfo {author} {\bibfnamefont {A.}~\bibnamefont
  {Accardi}} \emph {et~al.},\ }\bibfield  {title} {\enquote {\bibinfo {title}
  {{Strong Interaction Physics at the Luminosity Frontier with 22 GeV Electrons
  at Jefferson Lab}},}\ }\href@noop {} {\  (\bibinfo {year} {2023})},\ \Eprint
  {http://arxiv.org/abs/2306.09360} {arXiv:2306.09360 [nucl-ex]} \BibitemShut
  {NoStop}%
\bibitem [{\citenamefont {Bylinkin}\ \emph {et~al.}(2023)\citenamefont
  {Bylinkin} \emph {et~al.}}]{Bylinkin:2022rxd}%
  \BibitemOpen
  \bibfield  {author} {\bibinfo {author} {\bibfnamefont {A.}~\bibnamefont
  {Bylinkin}} \emph {et~al.},\ }\bibfield  {title} {\enquote {\bibinfo {title}
  {{Detector requirements and simulation results for the EIC exclusive,
  diffractive and tagging physics program using the ECCE detector concept}},}\
  }\href {\doibase 10.1016/j.nima.2023.168238} {\bibfield  {journal} {\bibinfo
  {journal} {Nucl. Instrum. Meth. A}\ }\textbf {\bibinfo {volume} {1052}},\
  \bibinfo {pages} {168238} (\bibinfo {year} {2023})},\ \Eprint
  {http://arxiv.org/abs/2208.14575} {arXiv:2208.14575 [physics.ins-det]}
  \BibitemShut {NoStop}%
\end{thebibliography}%
\end{document}